\documentclass[final,3p,times,onecolumn,sort&compress,hidelinks]{elsarticle}

\usepackage{graphicx}
\usepackage{amsmath}
\usepackage{amssymb}
\usepackage{mathtools}
\usepackage[utf8]{inputenc}
\usepackage[bookmarksopen,bookmarksnumbered]{hyperref}
\usepackage{color}
\usepackage[usenames,dvipsnames]{xcolor}
\usepackage[normalem]{ulem}
\usepackage{soul}
\usepackage{units}
\usepackage{rotating}
\usepackage{enumerate}
\usepackage{appendix}

\allowdisplaybreaks

\newcommand{\T}[1]{\boldsymbol{#1}_{\text{T}}}
\newcommand{\Tsc}[1]{#1_{\text{T}}}

\definecolor{dpmagenta}{rgb}{0.8, 0.0, 0.8}

\newcommand{\ba}{\begin{array}}
\newcommand{\ea}{\end{array}}
\newcommand{\beq}{\begin{equation}}
\newcommand{\eeq}{\end{equation}}

\newcommand\bstarsc{b_*}

\newcommand\mubstar{\mu_{\bstarsc}}

\newcommand{\arrowcom}[1]{\textcolor{red}{\textbf{$\Longrightarrow$ #1}} \\}

\def\??#1{\mbox{}\\\arrowcom{#1}}

\begin{document}

\begin{frontmatter}

\author[label1]{Leonard Gamberg}\ead{lpg10@psu.edu}
\author[label2]{Andreas Metz}\ead{metza@temple.edu}
\author[label1]{Daniel Pitonyak}\ead{dap67@psu.edu}
\author[label1,label3]{Alexei Prokudin} \ead{prokudin@jlab.org}
\address[label1]{Division of Science, Penn State University Berks, Reading, Pennsylvania 19610, USA}
\address[label2]{Department of Physics, SERC, Temple University, Philadelphia, PA 19122, USA}
\address[label3]{Theory Center, Jefferson Lab, 12000 Jefferson Avenue, Newport News, Virginia 23606, USA}

\title{Connections between collinear and transverse-momentum-dependent polarized observables within the Collins-Soper-Sterman formalism}

\begin{abstract}
 We extend the improved Collins-Soper-Sterman (iCSS) $W+Y$ construction recently presented in~\cite{Collins:2016hqq} to the case of polarized observables, where we focus in particular on the Sivers effect in semi-inclusive deep-inelastic scattering.  
 We further show how one recovers the expected leading-order collinear twist-3 result from a (weighted) $q_T$-integral of the differential cross section. We are also able to demonstrate the validity of the well-known relation between the (TMD) Sivers function and the (collinear twist-3) Qiu-Sterman function within the iCSS framework.  This relation allows for their interpretation as functions yielding the average transverse momentum of unpolarized quarks in a transversely polarized spin-$\frac{1}{2}$ target.  We further outline how this study can be generalized to other polarized quantities.
\end{abstract}


\end{frontmatter}

\date{\today}

\section{Introduction \label{sec:intro}}
One of the primary goals of transverse-momentum-dependent (TMD) factorization theorems~\cite{Collins:1981uk,Collins:1981uw, Collins:1984kg,Idilbi:2004vb,Collins:2004nx,Ji:2004wu,Collins:2011zzd,GarciaEchevarria:2011rb,Echevarria:2015usa}, which rely largely on the work of Collins, Soper, and Sterman (CSS)~\cite{Collins:1981uk,Collins:1981uw, Collins:1984kg},
is to describe the cross section as a function of the
transverse momentum $q_T = |\T{q}|$ point-by-point, from small $q_T\sim m$ (where $m$ is a typical hadronic mass scale), to large $q_T \sim Q$ (where $Q$ is a large momentum or mass in the reaction and sets the hard scale). 
 In order to achieve this, CSS organized the cross section in an additive form, $W+Y$.  The $W$-term is valid for $q_T\ll Q$ and involves TMD parton distribution functions (PDFs) and fragmentation functions (FFs).  The $Y$-term, which involves collinear PDFs and FFs, serves as a correction for larger $q_T$ values and is the difference between the collinear calculation
for $q_T\sim Q$ beginning at a fixed order in the strong coupling 
and its small transverse momentum asymptotic limit for $m\ll q_T\ll Q$~\cite{Collins:1984kg,Nadolsky:1999kb,Koike:2006fn}.

In recent years, much theoretical and phenomenological attention has been devoted
to the evolution of the  $W$-term expressed through TMD PDFs and/or FFs.
However, current studies~\cite{Sun:2013hua,Boglione:2014oea,Boer:2015uqa,Boglione:2016bph,Collins:2016hqq}
 indicate that a satisfactory treatment of non-zero
$\Tsc{q}/Q$ corrections and matching to the fixed-order ($\Tsc{q}\sim Q$) term is central to obtaining a leading-power approximation to the cross section over the whole range of $\Tsc{q}$.  In addition,  since collinear factorization is valid
not only for large $\Tsc{q}$, but also for the $\Tsc{q}$-integrated cross section, 
one expects to recover the collinear factorized result after integrating the TMD factorized differential 
cross section over $\T{q}$. 
The original CSS formulation did not address whether this connection is satisfied~\cite{Collins:2016hqq}.  While the matching of powers of $\Tsc{q}/Q$ in the intermediate $q_T$-region has been studied for semi-inclusive deep inelastic scattering (SIDIS) for various
polarizations~\cite{Ji:2006br,Bacchetta:2008xw}, in Refs.~\cite{Collins:2016hqq,Su:2014wpa,Boglione:2014oea} it was found that the original CSS $W+Y$ construction has difficulties to match the large- and small-$\Tsc{q}$ regions for phenomenological
studies at low $Q$.  
In addition, in Ref.~\cite{Collins:2016hqq} it was also
demonstrated that the original $W+Y$ construction does not
properly match collinear factorization for the cross section integrated over $\T{q}$.  

These issues led the authors of Ref.~\cite{Collins:2016hqq} to develop an improved version of the original CSS $W+Y$ construction (which we subsequently refer to as iCSS). The main focus of that work was for the case of the unpolarized SIDIS cross section, and it was demonstrated that for the iCSS construction the integral of $W+Y$ over $\T{q}$ reproduces the collinear result at leading order.  We also mention that the relation between resummed and collinear unpolarized results was discussed in Ref.~\cite{Voglesang:talkINT14} in the framework of joint resummation~\cite{Kulesza:2003wn}. 

With the intense focus on the 3D structure of hadrons through
spin-dependent observables, it is also important to extend the iCSS formalism to the polarized case, especially transverse single-spin asymmetries (TSSAs) like the Sivers~\cite{Sivers:1989cc,Sivers:1990fh} and Collins~\cite{Collins:1992kk} effects.  The purpose of this work is to achieve this goal.

We organize the Letter as follows: In Sec.~\ref{s:CSSrev} we review TMD evolution as it relates to the original CSS $W+Y$ construction along with issues that arise in both $\Tsc{q}$-matching in the intermediate region and matching the $q_T$-integrated cross section to
collinear factorization. Next, in Sec.~\ref{s:iCSS} we discuss the iCSS $W+Y$ method developed in Ref.~\cite{Collins:2016hqq} and
show how one can extend it to polarized observables, where in particular we focus on the Sivers effect in SIDIS.  Much data exists for this observable (see for instance Refs.~\cite{Airapetian:2009ae,Alekseev:2008dn,Adolph:2012sp,Qian:2011py}), including a recent measurement of the weighted asymmetry~\cite{Bradamante:2017yia}.  We also revisit the well-known relation between the (TMD) Sivers function and a (collinear) 3-parton correlator, the so-called Qiu-Sterman function~\cite{Efremov:1981sh,Efremov:1984ip,Qiu:1991pp,Qiu:1991wg}, in the context of the iCSS formalism.  In Sec.~\ref{s:physical} we discuss the importance of these results with regard to the interpretation of TMDs.  Finally, in Sec.~\ref{s:sum} we summarize our work.

\section{Review of the original CSS formalism \label{s:CSSrev}}

We begin with a synopsis of the $W+Y$ construction of the SIDIS $q_T$-differential cross section, which is given by 
\begin{equation} \label{e:SIDIS_WY}
 \Gamma(\T{q},Q,S) \equiv \frac{d\sigma} {dxdyd\phi_sdzd\phi_h(z^2dq_T^2)} = W(\T{q},Q,S)+Y(\T{q},Q,S)+O((m/Q)^{c}) \,,
\end{equation}
where $\T{q}$ and $-Q^2$ are the transverse momentum and virtuality, respectively, of the virtual photon, $S$ is a 4-vector for the spin of the target and $\phi_S$ the azimuthal angle of its transverse component, $\phi_h$ is the azimuthal angle of the produced hadron, and $x,\,y,\,z$ are the other standard SIDIS kinematic variables (see Ref.~\cite{Bacchetta:2006tn} for more details).  Note that in the arguments of $\Gamma$, $W$, and $Y$ we have suppressed the $x$ and $z$ dependence for brevity.  Also, we mention that the cross section oftentimes is written differential in the hadron transverse momentum $\boldsymbol{P}_{h\perp}=-z\T{q}$.  
In Eq.~(\ref{e:SIDIS_WY}), the $W$-term factorizes into TMD PDFs and FFs and is valid for $q_T\ll Q$, while the $Y$-term serves as a correction for larger $q_T$ values and uses collinear PDFs and FFs.  There is also a hard factor included in both terms.

The construction of the cross section in Eq.~(\ref{e:SIDIS_WY}) 
as the sum of $W(\T{q},Q,S)$ and $Y(\T{q},Q,S)$ 
results from applying so-called ``approximators'' to $\Gamma(\T{q},Q,S)$~\cite{Collins:2011zzd,Collins:2016hqq} 
that are designed to be valid for a certain region of $q_T$.
The  resulting cross section is accurate up to an error that is of order $(m/Q)^c$, where $c$ is a positive power, and $m$ is a typical hadronic mass scale.  The TMD approximator ${\rm T}_{\rm TMD}$ is valid for $q_T\ll Q$, while the collinear approximator ${\rm T}_{\rm coll}$ is valid for $q_T\sim Q$. 
Then one has
 $W(\T{q},Q,S)\equiv {\rm T}_{\rm TMD}\Gamma(\T{q},Q,S)$ and $Y(\T{q},Q,S) \equiv {\rm FO}(\T{q},Q,S)-{\rm AY}(\T{q},Q,S)$, where
${\rm FO}(\T{q},Q,S)\equiv {\rm T}_{\rm coll}\Gamma(\T{q},Q,S)$ is the fixed-order term and ${\rm AY}(\T{q},Q,S)\equiv {\rm T}_{\rm coll}{\rm T}_{\rm TMD}\Gamma(\T{q},Q,S)$ is the asymptotic term.  We note that the actual value for $c$ in the error term $O((m/Q)^{c})$ depends on which structure function we look at in $\Gamma(\T{q},Q,S)$.

\subsection{TMD evolution in coordinate space \label{ss:fact}}

In the CSS factorization formalism, the TMD evolution of the $W$-term in (\ref{e:SIDIS_WY}) is carried out in $b$-space.\footnote{A recent work performing the evolution in momentum space can be found in Ref.~\cite{Kang:2017cjk}.}  Thus, we focus on $\tilde{W}(\T{b},Q,S)$ and write $W(\T{q},Q,S)$ as its Fourier transform,
\begin{equation} \label{e:Wold}
W(\T{q},Q,S) = \int\!\frac{d^2\T{b}} {(2\pi)^2}\,e^{i\T{q}\cdot\T{b}}\,\tilde{W}(\T{b},Q,S)\,,
\end{equation}
where $\tilde{W}(\T{b},Q,S)$ can be expanded in the following structures~\cite{Boer:2011xd,Aybat:2011ge},
\begin{equation} \label{e:Wb_old}
  \tilde{W}(\T{b},Q,S) = \tilde{W}_{\rm UU}(b_T,Q) - iM_P\,\epsilon^{ij}b_T^iS_T^j\,\tilde{W}_{\rm UT}^{\rm siv}(b_T,Q) + \dots\;,
\end{equation}
with $M_P$ the mass of the target and the epsilon tensor defined with $\epsilon^{12} = 1$.  For our purposes, in $ \tilde{W}(\T{b},Q,S)$ we will focus only on the unpolarized and Sivers contributions, whereas the ellipsis indicate other azimuthal modulations that we will not address here. Note that because of the $b_T^i$ factor in the second term, $\tilde{W}_{\rm UT}^{\rm siv}(b_T,Q)$ does not have a kinematic zero at $b_T=0$.  From (\ref{e:Wb_old}) it immediately follows that
\begin{equation}
W(\T{q},Q,S) = W_{\rm UU}(q_T,Q)+|\T{S}|\sin(\phi_h-\phi_S)\,W_{\rm UT}^{\rm siv}(q_T,Q) + \dots,
\label{e:WqT}
\end{equation}
where the Fourier transforms take the form
\begin{align}
  W_{\rm UU}(q_T,Q) &\equiv \int\!\frac{d^2\T{b}}{(2\pi)^2}\,e^{i\T{q}\cdot\T{b}}\,\tilde{W}_{\rm UU}(b_T,Q)=\int_0^\infty \frac{db_T}{2\pi}\, b_T J_0(q_Tb_T)\tilde{W}_{\rm UU}(b_T,Q)  \,,\label{e:WUU_q_old}\\[0.3 cm]
  W_{\rm UT}^{\rm siv}(q_T,Q) &\equiv -iM_P\int\!\frac{d^2\T{b}} {(2\pi)^2}\,e^{i\T{q}\cdot\T{b}}\,(\boldsymbol{\hat{h}}\cdot\T{b})\,\tilde{W}_{\rm UT}^{\rm siv}(b_T,Q)=-M_p\int_0^\infty \frac{db_T}{2\pi}b_T^2J_1(q_Tb_T)\tilde{W}_{\rm UT}^{\rm siv}(b_T,Q)\,,
  \label{e:WUT_q_old}
\end{align}
where $\boldsymbol{\hat{h}} = \boldsymbol{P_{h\perp}}/P_{h\perp}=-\T{q}/q_T$.  Note that $W_{\rm UT}^{\rm siv}(q_T,Q)$ has a kinematic zero at $q_T=0$.

The scalar functions in (\ref{e:Wb_old}) can be expressed in terms of Fourier transformed TMDs (FT-TMDs)~\cite{Collins:1981uk,Collins:1981uw,Collins:1984kg,Collins:2011zzd,Aybat:2011zv,Boer:2011xd,Aybat:2011ge}.
The unpolarized scalar function is
\begin{align}
  \tilde{W}_{\rm UU}(b_T,Q)& = \sum_j H_j(\mu,Q)\,
  \tilde{f}_1^{j}(x,b_T;\zeta_A,\mu)\,
  \tilde{D}_1^{h/j}(z,b_T;\zeta_B,\mu)\,,\label{e:WbUU_old0}
  \end{align}
where $\tilde{f}_1^{j}\big( x, b_T  ; \zeta_A,\mu \bigr)$
and $\tilde{D}_1^{h/j} \big( z, b_T ; \zeta_B,\mu \bigr)$ are, 
respectively, the unpolarized
FT-TMD PDF and FF.
The FT-TMDs have two scale arguments: $\mu$, which is the renormalization scale, and $\zeta$, which parameterizes how the effects of soft-gluon radiation are partitioned between the FT-TMDs, where $\zeta_A\zeta_B=Q^4$.  We use the freedom from the renormalization group to set $\zeta_A=\zeta_B=Q^2$ and $\mu=C_2Q\equiv \mu_Q$.  The constant $C_2$ is chosen to optimize the accuracy of the perturbation theory for the lepton-quark hard scattering coefficient $H_j(\mu,Q)$, which at LO is $H^{{\rm LO}}_{j}(\mu_Q,Q)=\alpha_{em}^2\,e_j^2(1-y+y^2/2)/(yQ^2)$.
With these choices, the unpolarized scalar function now reads
\begin{equation}
  \tilde{W}_{\rm UU}(b_T,Q) = \sum_j H_j(\mu_Q,Q)\,\tilde{f}_1^{j}(x,b_T;Q^2,\mu_Q)\,\tilde{D}_1^{h/j}(z,b_T;Q^2,\mu_Q)\,.\label{e:WbUU_old}
 \end{equation}
Similarly, the Sivers scalar function reads
\begin{equation}
  \tilde{W}_{\rm UT}^{\rm siv}(b_T,Q) = \sum_j H_j(\mu_Q,Q)\,\tilde{f}_{1T}^{\perp(1)j}(x,b_T;Q^2,\mu_Q)\,\tilde{D}_1^{h/j}(z,b_T;Q^2,\mu_Q)\,,\label{e:WbUT_old}
\end{equation}
where~\cite{Boer:2011xd,Aybat:2011ge} 
 \begin{equation} \label{e:Siv_mom_bT}
 \tilde{f}_{1T}^{\perp(1)j}(x,b_T;Q^2,\mu_Q)\equiv -\frac{1} {M_P^2\,b_T} \frac{\partial\tilde{f}_{1T}^{\perp j}(x,b_T;Q^2,\mu_Q)} {\partial b_T}\,,
 \end{equation}
and $\tilde{f}_{1T}^{\perp j}(x,b_T;Q^2,\mu_Q)$ is the FT-TMD Sivers function.\footnote{In the limit $b_T \to 0$, one finds $\tilde{f}_{1T}^{\perp(1)}(x,b_T\to 0) = \int d^2\T{k} \frac{k_T^2} {2M_P^2}\,f_{1T}^{\perp}(x,k_T) \equiv f_{1T}^{\perp(1)}(x)$ (where $k_T = |\T{k}|$)~\cite{Boer:2011xd}.}  The expressions in Eqs.~(\ref{e:WbUU_old}), (\ref{e:WbUT_old}) lead to (after using (\ref{e:WUU_q_old}), (\ref{e:WUT_q_old})) the well-known results in momentum-space for $W_{\rm UU}(q_T,Q)$ and $W_{\rm UT}^{\rm siv}(q_T,Q)$ in Eq.~(\ref{e:WqT})~\cite{Aybat:2011zv,Collins:2011zzd},
\begin{align}
W_{\rm UU}(q_T,Q) &=  \sum_j H_j(\mu_Q,Q)\int \!d^2\T{k}\,d^2\T{p}\,\delta^{(2)}(\T{k}-\T{p}+\T{q})\,f_1^j(x,k_T;Q^2,\mu_Q)\,D_1^{h/j}(z,zp_T;Q^2,\mu_Q)\,,\\
W_{\rm UT}^{\rm siv}(q_T,Q) &= \sum_j H_j(\mu_Q,Q)\int \!d^2\T{k}\,d^2\T{p}\,\delta^{(2)}(\T{k}-\T{p}+\T{q})\,\left(-\frac{\boldsymbol{\hat{h}}\cdot\T{k}} {M_P}\right)f_{1T}^{\perp\,j}(x,k_T;Q^2,\mu_Q)\,D_1^{h/j}(z,zp_T;Q^2,\mu_Q)\,,
\end{align}
which in their structure agree with the parton model calculations~\cite{Mulders:1995dh,Bacchetta:2006tn,Boer:1997nt}.  The variable $\T{k}$ is the transverse momentum of the struck quark w.r.t.~the incoming proton, and $\T{p}$ is transverse momentum of the fragmenting quark w.r.t.~the produced hadron (with $k_T=|\T{k}|$, $p_T=|\T{p}|$).

We mention that the FT-TMDs or ``$b$-space functions'' on the r.h.s.~of (\ref{e:WbUU_old}), (\ref{e:WbUT_old}) can be understood simply as functions arising from the factorization of $\tilde{W}_{\rm UU}(b_T,Q)$ and $\tilde{W}^{\rm siv}_{\rm UT}(b_T,Q)$ from which we can {\it define} the momentum-space functions through inverse Fourier transforms.  Likewise, $W(\T{q},Q,S)$ can be
{\it defined} from $\tilde{W}(\T{b},Q,S)$ via the inverse Fourier transform~(\ref{e:Wold}).  We will adopt this viewpoint, rather than the usual approach of calculating $b$-space functions from the Fourier transform of the momentum-space correlator (see, e.g., Ref.~\cite{Boer:2011xd} for details), since TMD evolution and certain modifications in the iCSS formalism are performed directly in $b$-space~\cite{Collins:2016hqq,Collins:2011zzd} (see Sec.~\ref{s:iCSS}). 

We next follow the CSS procedure~\cite{Collins:2011zzd,Collins:2014jpa} to write Eqs.~(\ref{e:WbUU_old}) and (\ref{e:WbUT_old}), respectively, as 
\begin{subequations}
\begin{align}
\tilde{W}_{\rm UU}(b_T,Q)&= \tilde{W}_{\rm UU}^{\rm OPE}(b_*(b_T),Q)\tilde{W}_{\rm UU}^{\rm NP}(b_T,Q)\\[0.3cm] 
& = \sum_j H_j(\mu_Q,Q)\,\tilde{f}_1^{j}(x,b_*(b_T);\mu_{b_*}^2,\mu_{b_*})\,\tilde{D}_1^{h/j}(z,b_*(b_T);\mu_{b_*}^2,\mu_{b_*})\nonumber\\
&\times\exp\left\{\tilde{K}(b_*(b_T);\mu_{b_*})\ln\left(\frac{Q^2} {\mu_{b_*}^2}\right)+\int_{\mu_{b_*}}^{\mu_Q}\!\frac{d\mu^\prime} {\mu^\prime}\left[2\gamma(\alpha_s(\mu^\prime);1)-\ln\left(\frac{Q^2} {\mu^{\prime 2}}\right)\gamma_K(\alpha_s(\mu^\prime))\right]\right\}\nonumber\\
&\times\exp\left\{-g_{\rm pdf}(x,b_T;Q_0,b_{max})-g_{\rm ff}(z,b_T;Q_0,b_{max})-g_K(b_T;b_{max})\ln\left(\frac{Q^2}{Q_0^2}\right)\right\}\,,\label{e:WbUUcss_old}
\end{align}
\end{subequations}
\begin{subequations}
\begin{align}
\tilde{W}_{\rm UT}^{\rm siv}(b_T,Q)&= \tilde{W}_{\rm UT}^{\rm siv, OPE}(b_*(b_T),Q)\tilde{W}_{\rm UT}^{\rm siv, NP}(b_T,Q)\\[0.3cm]
& = \sum_j H_j(\mu_Q,Q)\,\tilde{f}_{1T}^{\perp(1)j}(x,b_*(b_T);\mu_{b_*}^2,\mu_{b_*})\,\tilde{D}_1^{h/j}(z,b_*(b_T);\mu_{b_*}^2,\mu_{b_*})\nonumber\\
&\times\exp\left\{\tilde{K}(b_*(b_T);\bar{\mu})\ln\left(\frac{Q^2} {\mu_{b_*}^2}\right)+\int_{\mu_{b_*}}^{\mu_Q}\!\frac{d\mu^\prime} {\mu^\prime}\left[2\gamma(\alpha_s(\mu^\prime);1)-\ln\left(\frac{Q^2} {\mu^{\prime 2}}\right)\gamma_K(\alpha_s(\mu^\prime))\right]\right\}\nonumber\\
&\times\exp\left\{-g_{\rm siv}(x,b_T;Q_0,b_{max})-g_{\rm ff}(z,b_T;Q_0,b_{max})-g_K(b_T;b_{max})\ln\left(\frac{Q^2}{Q_0^2}\right)\right\}\,,\label{e:WbUTcss_old}
\end{align}
\end{subequations}
with
\begin{equation}
b_*(b_T) \equiv \sqrt{\frac{b_T^2} {1+b_T^2/b_{max}^2}}\,,\quad\quad \mu_{b_*} \equiv \frac{C_1} {b_*(b_T)}\,,
\end{equation}
where $b_{max}$ separates small and large $b_T$, and $C_1$ is a constant chosen to allow for perturbative calculations of $b_*(b_T)$-dependent quantities
without large logarithms~\cite{Collins:2014jpa}.  Note that $b_*(b_T)$ freezes at $b_{max}$ when $b_T$ is large so that $b_*(b_T)$ is always small (i.e., in the perturbative region).  The first two lines in (\ref{e:WbUUcss_old}), (\ref{e:WbUTcss_old}) are the operator product expansion (OPE) pieces, $\tilde{W}^{\rm OPE}(b_*(b_T),Q)$, whereas the last line is the non-perturbative part, $\tilde{W}^{\rm NP}(b_T,Q)$.
The functions $g_{\rm pdf}$ and $g_{\rm ff}$ are the non-perturbative factors for $\tilde{f}_1$ and $\tilde{D}_1$, respectively, while $g_{\rm siv}$ is the non-perturbative factor for $\tilde{f}_{1T}^{\perp(1)}$.  The factor $g_K$ is the non-perturbative part of the Collins-Soper (CS) evolution kernel $\tilde{K}(b_T;\mu)$ (see~\cite[Eqs.~(6), (11), (25)]{Collins:2014jpa}).  Note that $\tilde{W}^{\rm NP} \to 1$ as $b_T \to 0$~\cite{Collins:2011zzd,Collins:2014jpa}.  
The terms $\gamma_K(\alpha_s(\mu))$ and $\gamma(\alpha_s(\mu); 1)$ are
the anomalous dimensions for the CS kernel and $b$-space functions, respectively
(see~\cite[Eqs.~(7)--(10), (12)]{Collins:2014jpa}).\footnote{See also~\cite{Rogers:2015sqa,Collins:2012ss} and references therein for
detailed discussions of the evolution equations and their origins.}  

We mention that there are some alternatives in the literature to the $b_*$-prescription. 
In Refs.~\cite{Qiu:2000ga,Qiu:2000hf} the authors separate the perturbative and non-perturbative contribution through the parameter $b_{max}$ such that $\tilde{W}(b_T,Q)=\tilde{W}(b_T,Q) \;{\rm for}\; b_T \leq b_{max}$ and $\tilde{W}(b_T,Q)=\tilde{W}(b_{max},Q)\,\tilde{W}^{\rm NP}_{QZ}(b_T,Q;b_{max})\;{\rm for}\;b_T > b_{max}$, where $\tilde{W}^{\rm NP}_{QZ}(b_T,Q;b_{max})$ includes power corrections to improve the matching between the perturbative and non-perturbative regions of $\tilde{W}(b_T,Q)$.
Their approach  attempts to minimize
the influence of the non-perturbative piece of $\tilde{W}(b_T,Q)$, which contains several parameters and does not have a fixed functional form, at small $b_T$ where one should be driven by perturbatively calculable effects.  In the context of the ``resummation approach''~\cite{Laenen:2000de,Kulesza:2002rh}, 
one avoids the  Landau pole encountered
in  performing Fourier transforms ($b$-space integrations)
by extending $\Tsc{b}$ to the complex plane
and exploiting the analytic structure of the running coupling.  Phenomenological parameters then appear only as non-perturbative power corrections.
In this work we  continue to use the CSS $b_*$-prescription.

Since the unpolarized $b$-space functions
in the first line of Eq.~(\ref{e:WbUUcss_old})
are restricted to small $b_T$, we can expand them in an OPE in terms of twist-2 collinear functions~\cite{Collins:2011zzd,Aybat:2011zv,Collins:2014jpa},
\begin{align}
\tilde{f}_1^{j}(x,b_*(b_T);\mu_{b_*}^2,\mu_{b_*}) &= \sum_{j'}\int_x^1\!\frac{d\hat{x}} {\hat{x}}\,\tilde{C}^{\rm pdf}_{j/j'}(x/\hat{x},b_*(b_T);\mu_{b_*}^2,\mu_{b_*},\alpha_s(\mu_{b_*}))\,f_1^{j'}(\hat{x};\mu_{b_*}) + O((m\,b_*(b_T))^p)\,,\label{e:OPE_f1_old}\\[0.3cm]
\tilde{D}_1^{h/j}(z,b_*(b_T);\mu_{b_*}^2,\mu_{b_*}) &= \sum_{i'}\int_z^1\!\frac{d\hat{z}} {\hat{z}^3}\,\tilde{C}^{\rm ff}_{i'/j}(z/\hat{z},b_*(b_T);\mu_{b_*}^2,\mu_{b_*},\alpha_s(\mu_{b_*}))\,D_1^{h/i'}(\hat{z};\mu_{b_*}) + O((m\,b_*(b_T))^p)\, \label{e:OPE_D1_old},
\end{align}
where $f_1^{j'}(\hat{x};\mu_{b_*})$, $D_1^{h/i'}(\hat{z};\mu_{b_*})$ 
 showing up on the r.h.s.~of Eqs.~(\ref{e:OPE_f1_old}), (\ref{e:OPE_D1_old}), respectively, are understood to be renormalized at the scale $\mu_{b_*}$. 

Similarly, the $b$-space function in Eq.~(\ref{e:WbUT_old}) that gives rise to the Sivers effect can be written at small $b_T$ in terms of a twist-3 quark-gluon-quark correlation function~\cite{Ji:2006ub,Ji:2006br,Kang:2011mr,Aybat:2011ge}, 
\begin{align}
\tilde{f}_{1T}^{\perp(1)j}(x,b_*(b_T);\mu_{b_*}^2,\mu_{b_*}) &= -\frac{1} {2M_P}\!\sum_{j'}\!\int_x^1\!\frac{d\hat{x}_1} {\hat{x}_1}\frac{d\hat{x}_2} {\hat{x}_2}\tilde{C}^{\rm siv}_{j/j'}(\hat{x}_1,\hat{x}_2,b_*(b_T);\mu_{b_*}^2,\mu_{b_*},\alpha_s(\mu_{b_*}))T_F^{j'}(\hat{x}_1,\hat{x}_2;\mu_{b_*}) \nonumber\\[-0.2cm]
&\hspace{9cm}+ O((m\,b_*(b_T))^{p'})\,, \label{e:OPE_Siv_old}
\end{align}
where $T_F^{j'}(\hat{x}_1,\hat{x}_2;\mu_{b_*})$~\footnote{Terms in Eq.~(\ref{e:OPE_Siv_old}) proportional to the derivative of $T_F(x_1,x_2)$ can be transformed into non-derivative pieces through an integration by parts, leading to a generalized coefficient function in (\ref{e:OPE_Siv_old}) (see, e.g., Ref.~\cite{Zhou:2009jm}).}
likewise is also  understood to be renormalized at the scale $\mu_{b_*}$. 
The function $T_F(x_1,x_2)$ in (\ref{e:OPE_Siv_old}) has the following
 operator definition:
\begin{equation} \label{e:T_F}
T_F(x_1,x_2) = \int\!\frac{db^- dy^-} {4\pi}\,e^{ix_1P^+b^-}e^{i(x_2-x_1)P^+y^-}\,\epsilon^{ij}S_T^j\,\langle P,S|\bar{\psi}(0)\gamma^+\mathcal{W}(0;y^-)gF^{+i}(y^-)\mathcal{W}(y^-;b^-)\psi(b^-)|P,S\rangle\,,
\end{equation}
where $\mathcal{W}(a^-;b^-)$ is a straight-line gauge link connecting $(0^+,a^-,\T{0})$ to $(0^+,b^-,\T{0})$.  When $x_1=x_2\equiv x$, $T_F(x,x)$ is known as the Qiu-Sterman function~\cite{Efremov:1981sh,Efremov:1984ip,Qiu:1991pp,Qiu:1991wg}.  Note that the gauge link for $\tilde{f}_{1T}^{\perp(1)j}(x,b_T)$ is understood to be for SIDIS.  If instead one used a gauge link consistent with the Drell-Yan process, the sign on the r.h.s.~of (\ref{e:OPE_Siv_old}) would change~\cite{Collins:2002kn}.  The errors in (\ref{e:OPE_f1_old})--(\ref{e:OPE_Siv_old}) are suppressed by positive powers $p,p'$, and represent higher twist terms.   

Note that the unpolarized $b$-space functions in (\ref{e:OPE_f1_old}), (\ref{e:OPE_D1_old}) are written at small $b_T$ (or large $k_T$ in momentum space) in terms  of twist-2 functions while the $b$-space function in (\ref{e:OPE_Siv_old}) associated with the Sivers effect is written in terms of a twist-3 function.  The reason is due to their different power-law behaviors at large-$k_T$: the former goes as $1/k_T^2$ while the latter is suppressed by a power of $k_T$ and goes as $1/k_T^3$~\cite{Ji:2006ub,Ji:2006br,Kang:2011mr,Aybat:2011ge,Bacchetta:2008xw}.

Using Eqs.~(\ref{e:OPE_f1_old})--(\ref{e:OPE_Siv_old}) in (\ref{e:WbUUcss_old}), (\ref{e:WbUTcss_old}) and comparing the results to Eqs.~(\ref{e:WbUU_old}), (\ref{e:WbUT_old}) allows us to establish the following equalities:
\begin{align}
\tilde{f}_1^{j}(x,b_T;Q^2,\mu_Q) &= \sum_{j'}\int_x^1\!\frac{d\hat{x}} {\hat{x}}\,\tilde{C}^{\rm pdf}_{j/j'}(x/\hat{x},b_*(b_T);\mu_{b_*}^2,\mu_{b_*},\alpha_s(\mu_{b_*}))\,f_1^{j'}(\hat{x};\mu_{b_*})\nonumber\\
&\times\exp\left\{\tilde{K}(b_*(b_T);\mu_{b_*})\ln\left(\frac{Q} {\mu_{b_*}}\right)+\int_{\mu_{b_*}}^{\mu_Q}\!\frac{d\mu^\prime} {\mu^\prime}\left[\gamma(\alpha_s(\mu^\prime);1)-\ln\left(\frac{Q} {\mu^{\prime}}\right)\gamma_K(\alpha_s(\mu^\prime))\right]\right\}\nonumber\\
&\times\exp\left\{-g_{\rm pdf}(x,b_T;Q_0,b_{max})-g_K(b_T;b_{max})\ln\left(\frac{Q}{Q_0}\right)\right\}\,,\label{e:f1_old}\\[0.3cm]
\tilde{D}_1^{h/j}(z,b_T;Q^2,\mu_Q) &= \sum_{i'}\int_z^1\!\frac{d\hat{z}} {\hat{z}^3}\,\tilde{C}^{\rm ff}_{i'/j}(z/\hat{z},b_*(b_T);\mu_{b_*}^2,\mu_{b_*},\alpha_s(\mu_{b_*}))\,D_1^{h/i'}(\hat{z};\mu_{b_*})\nonumber\\
&\times\exp\left\{\tilde{K}(b_*(b_T);\mu_{b_*})\ln\left(\frac{Q} {\mu_{b_*}}\right)+\int_{\mu_{b_*}}^{\mu_Q}\!\frac{d\mu^\prime} {\mu^\prime}\left[\gamma(\alpha_s(\mu^\prime);1)-\ln\left(\frac{Q} {\mu^{\prime}}\right)\gamma_K(\alpha_s(\mu^\prime))\right]\right\}\nonumber\\
&\times\exp\left\{-g_{\rm ff}(z,b_T;Q_0,b_{max})-g_K(b_T;b_{max})\ln\left(\frac{Q}{Q_0}\right)\right\}\,,\label{e:D1_old}\\[0.3cm]
\tilde{f}_{1T}^{\perp(1)j}(x,b_T;Q^2,\mu_Q) &= -\frac{1} {2M_P}\sum_{j'}\int_x^1\!\frac{d\hat{x}_1} {\hat{x}_1}\frac{d\hat{x}_2} {\hat{x}_2}\,\tilde{C}^{\rm siv}_{j/j'}(\hat{x}_1,\hat{x}_2,b_*(b_T);\mu_{b_*}^2,\mu_{b_*},\alpha_s(\mu_{b_*}))\,T_F^{j'}(\hat{x}_1,\hat{x}_2;\mu_{b_*})\nonumber\\
&\times\exp\left\{\tilde{K}(b_*(b_T);\mu_{b_*})\ln\left(\frac{Q} {\mu_{b_*}}\right)+\int_{\mu_{b_*}}^{\mu_Q}\!\frac{d\mu^\prime} {\mu^\prime}\left[\gamma(\alpha_s(\mu^\prime);1)-\ln\left(\frac{Q} {\mu^{\prime}}\right)\gamma_K(\alpha_s(\mu^\prime))\right]\right\}\nonumber\\
&\times\exp\left\{-g_{\rm siv}(x,b_T;Q_0,b_{max})-g_K(b_T;b_{max})\ln\left(\frac{Q}{Q_0}\right)\right\}\, ,\label{e:Siv_old}
\end{align}
where implicitly there are again errors in (\ref{e:f1_old})--(\ref{e:Siv_old}) due to the truncation of the OPE in~(\ref{e:OPE_f1_old})--(\ref{e:OPE_Siv_old}).
From these $b$-space expressions, we {\it define} the following functions in momentum space as
\begin{align}
  f_1^{j}(x,k_T;Q^2,\mu_Q)&\equiv \int_0^\infty \!\frac{db_T}{2\pi}\, b_T J_0(k_Tb_T)\tilde{f}_1^{j}(x,b_T;Q^2,\mu_Q)
  \, ,\label{e:f1kT_old}\\[0.3cm]
  D_1^{h/j}(z,zp_T;Q^2,\mu_Q)&\equiv \int_0^\infty \!\frac{db_T}{2\pi}\, b_T J_0(p_Tb_T)\,\tilde{D}_1^{h/j}(z,b_T;Q^2,\mu_Q)
  \, , \label{e:D1pT_old}\\[0.3cm]
 \frac{k_T^2} {2M_P^2}\,f_{1T}^{\perp j}(x,k_T;Q^2,\mu_Q)&\equiv k_T\!\!\int_0^\infty \!\frac{db_T}{4\pi}b_T^2J_1(k_Tb_T)
  \,\tilde{f}_{1T}^{\perp(1)j}(x,b_T;Q^2,\mu_Q)  \,,
  \label{e:SivkT_old}
\end{align} 
where $k_T = |\T{k}|$, $p_T = |\T{p}|$. The definitions in Eqs.~(\ref{e:f1kT_old})--(\ref{e:SivkT_old}) are consistent with the standard parameterization of the momentum-space (distribution and fragmentation) correlators~\cite{Mulders:1995dh,Goeke:2005hb,Boer:1997mf} in terms of (among others) the unpolarized TMDs $f_1(x,k_T)$, $ D_1(z,zp_T)$ and the Sivers TMD $f_{1T}^{\perp}(x,k_T)$.  In particular, the result in (\ref{e:SivkT_old}) can be obtained from Eq.~(\ref{e:Siv_mom_bT}) along with the explicit expression for the Fourier transform of the Sivers function.  The inverse Fourier transforms read
\begin{align}
\tilde{f}_1^{j}(x,b_T;Q^2,\mu_Q)  & = 2 \pi \int_0^\infty dk_T k_T J_0(k_Tb_T)\, f_1^{j}(x,k_T;Q^2,\mu_Q)
  \, ,\label{e:f1kT_inverse}\\[0.3cm]
 \tilde{D}_1^{h/j}(z,b_T;Q^2,\mu_Q) & =2 \pi \int_0^\infty dp_T p_T J_0(p_Tb_T)\,
 D_1^{h/j}(z,zp_T;Q^2,\mu_Q) \, , \label{e:D1pT_inverse}\\[0.3cm]
 \tilde{f}_{1T}^{\perp(1)j}(x,b_T;Q^2,\mu_Q)& = \frac{2 \pi} {M_P^2}\; \!\!\int_0^\infty \!dk_T \frac{k_T^2}{b_T}J_1(k_Tb_T)
  \, \,f_{1T}^{\perp j}(x,k_T;Q^2,\mu_Q) \,.
  \label{e:SivkT_inverse}
\end{align} 
We are now in a position to address some issues with the original CSS formalism presented above.

\subsection{Issues with the original $W+Y$ construction \label{ss:issues}}
We will highlight two main issues with the original CSS $W+Y$ construction that are also detailed in Ref.~\cite{Collins:2016hqq},  and serve as motivation for the modifications we will discuss in Sec.~\ref{s:iCSS}.  The first is 
that the CSS framework is most useful when $Q$ is large enough that there is a broad intermediate range of transverse momentum characterized by $m \ll \Tsc{q} \ll Q$. 
That is, one needs to have a window in $q_T$ where $\Tsc{q}/Q$ is
small enough that factorization using TMD PDFs and FFs is valid to sufficient accuracy~\cite{Collins:2016hqq,Collins:2011zzd}
while $m/\Tsc{q}$ is also small enough that
factorization using collinear PDFs and FFs is simultaneously valid. However, at the values of $Q$ that are of phenomenological interest, for example, in measurements devoted to studying 3D hadronic structure through the intrinsic transverse motion of partons, neither $\Tsc{q}/Q$ nor $m/\Tsc{q}$ is necessarily small.
Given these conditions,
it becomes a challenge to smoothly and consistently match the differential cross section over the available range of $q_T$~\cite{Su:2014wpa,Boglione:2014oea}.  

The second issue is the problem of matching the TMD factorized cross section
 integrated over $\T{q}$ to the collinear factorization formalism.
 As the authors of Ref.~\cite[Appendix A]{Collins:2016hqq} showed,
 the integral of $\Gamma(\T{q},Q,S)$ over all $\T{q}$ in Eq.~(\ref{e:SIDIS_WY})
 results in a mismatch of orders in $\alpha_S(Q)$ of the leading contributions
 on the l.h.s.~and r.h.s.~of the equation.
This is evident from the fact that integrating $W(\T{q},Q,S)$ over all $\T{q}$ gives zero instead of the expected collinear result.  To be more specific, integrating Eq.~(\ref{e:WqT}) over $\T{q}$ and using Eq.~(\ref{e:WUU_q_old}) yields~\cite{Collins:2016hqq}
\begin{align}
    \int\!d^2\T{q}\, W(\T{q},Q,S) &= \tilde{W}_{\rm UU}(b_T\to 0,Q)\nonumber\\
    &\sim b_T^{a}\times({\rm log\;corrections}) = 0 \,, \label{e:unp_parton_old}
\end{align}
where $a = 8C_F/\beta_0$ with $\beta_0 = 11-2n_f/3$.  (Note that the integration over $\T{q}$ eliminates all other terms in $W(\T{q},Q,S)$ except for $W_{\rm UU}(q_T,Q)$.) The source of this behavior is that $b_*(b_T\to 0) = 0$ so that $\mu_{b_*} \to \infty$ in this limit.  This leads to a large logarithm in the second term of the perturbative (OPE) exponential (second line of (\ref{e:WbUUcss_old})) involving $\gamma_K(\alpha_s(\mu_{b_*}))\ln(Q^2/\mu_{b_*})$~\cite{Collins:2016hqq}. Similarly, for the Sivers contribution to $W(\T{q},Q,S)$
we find, using Eqs.~(\ref{e:WqT}) and (\ref{e:WUT_q_old}),
\begin{align}
 \int\!d^2\T{q}\,q_{T}\sin(\phi_h-\phi_S)\, W(\T{q},Q,S)
&=\pi \!\int dq_T\, q_T^2\,W_{\rm UT}^{\rm Siv}(q_T,Q)\nonumber\\[0.15cm]
&=-M_P\lim_{b'_T\to 0}\frac{1}{b'_T}\left[\int_0^\infty\! db_T\,b_T\!\int_0^\infty\! dq_T\,q_Tb_T\,J_1(q_Tb'_T)\,J_1(q_Tb_T)\,\tilde{W}_{\rm UT}^{\rm siv}(b_T,Q)\right]\nonumber\\[0.15cm]
&=-M_P\,\tilde{W}^{\rm siv}_{\rm UT}(b_T\to 0,Q)\nonumber\\[0.15cm]
&\sim b_T^{a}\times({\rm log\;corrections}) = 0 \,,\label{e:Siv_parton_old}
\end{align}
with $a$ as above. The last line holds because the perturbative (OPE) part of the Sudakov exponential is independent of spin (cf.~Eqs.~(\ref{e:WbUUcss_old}) and (\ref{e:WbUTcss_old})), so $\tilde{W}^{\rm siv}_{\rm UT}(b_T,Q)$ retains the same behavior when $b_T\to 0$ as $\tilde{W}_{\rm UU}(b_T,Q)$.  In going from the second to the third line in Eq.~(\ref{e:Siv_parton_old}) we exploited the
well-known relation used in Bessel weighting~\cite{Boer:2011xd}, $\int_0^\infty\! dq_T\,q_TJ_n(q_Tb'_T)J_m(q_Tb_T)= \delta_{nm}\,\delta(b_T-b'_T)/b_T$.  Therefore, from (\ref{e:Siv_parton_old}) we see the weighted Sivers effect also vanishes instead of giving the expected collinear twist-3 expression.

From the above results, one can readily conclude that the integrals over $\T{k}$ (or $\T{p}$) of the unpolarized functions (\ref{e:f1kT_old}), (\ref{e:D1pT_old}) vanish upon integration over transverse momentum,
\begin{align}
\int\!d^2\T{k}\,f_1^{j}(x,k_T;Q^2,\mu_Q)&=\tilde{f}_1^{j}(x,b_T\to 0;Q^2,\mu_Q)=0 \label{e:f1_col_old}\,,\\[0.1cm]
z^2\int\!d^2\T{p}\,D_1^{j}(z,zp_T;Q^2,\mu_Q)&=z^2\tilde{D}_1^{h/j}(z,b_T\to 0;Q^2,\mu_Q)=0 \label{e:D1_col_old}\, ,
\end{align}
and likewise the first moment of the Sivers function vanishes,
\begin{align}
\int\!d^2\T{k}\,\frac{k_T^2} {2M_P^2}\,f_{1T}^{\perp j}(x,k_T;Q^2,\mu_Q)&\equiv f_{1T}^{\perp(1)\,j}(x;Q^2,\mu_Q) =\tilde{f}_{1T}^{\perp(1)j}(x,b_T\to 0;Q^2,\mu_Q)=0 \label{e:Siv_col_old}\,.
\end{align}
Note that a dramatic consequence of (\ref{e:f1_col_old})--(\ref{e:Siv_col_old}) is that the physical interpretation of integrated TMDs is lost.  For example, the far l.h.s.~of Eq.~(\ref{e:Siv_col_old}) is supposed to determine the average transverse momentum of unpolarized quarks in a transversely polarized spin-$\frac{1} {2}$ target~\cite{Boer:1997bw,Boer:1997nt,Burkardt:2003yg,Burkardt:2004ur,Meissner:2007rx}.  Clearly, such a statement is not true in the original CSS framework. In Ref.~\cite{Collins:2016hqq} this formalism is amended in order to address these issues for the unpolarized case.  In the next section we review these improvements and explain how they are implemented for the polarized case.  More importantly, we demonstrate in Sec.~\ref{s:physical} how one can restore the standard physical interpretation of (\ref{e:f1_col_old})--(\ref{e:Siv_col_old}) at leading order (LO).

\section{The improved CSS formalism \label{s:iCSS}}

\subsection{Modifications to the original CSS framework \label{ss:mod}}
In order to deal with some of the problems of the original CSS $W+Y$ construction discussed in the previous section, the authors of Ref.~\cite{Collins:2016hqq} incorporated improvements to the formalism.  We briefly summarize their prescriptions below (see Ref.~\cite{Collins:2016hqq} for more details), which originally were for the unpolarized SIDIS cross section, and extend their implementation to the case of the Sivers effect.  We will briefly outline at the end of this section how to generalize the iCSS procedure for any polarized observable.

In what follows, we discuss four steps from the iCSS formulation.  Step (I) addresses the vanishing of $\tilde{W}_{UU}(b_T,Q)$ and $\tilde{W}_{UT}^{\rm siv}(b_T,Q)$ at $b_T = 0$, so that one can have a factorized collinear expansion in terms of PDFs and FFs in this limit.  Steps (II) and (III) help improve the matching between the $W$-term and $Y$-term in the intermediate-$q_T$ regime by restricting them to their respective region of applicability
(see also Ref.~\cite{Catani:2015vma}).  Step (IV) collects these modifications to form the $q_T$-differential cross section $\Gamma(\T{q},Q,S)$.
\begin{enumerate}[(I)]
\item Replace $b_T$ with $b_c(b_T)$ in order to deal with the large logarithms that arise as $b_T\to 0$ (see also~\cite{Bozzi:2005wk,Parisi:1979se}), where
\begin{equation}  \label{e:bcbT}
b_c(b_T) = \sqrt{b_T^2+\left(\frac{b_0}{C_5 Q}\right)^{\!2}} = \sqrt{b_T^2+b_{min}^{\prime 2}}\,,
\end{equation}
with $b_0\equiv 2\exp(-\gamma_E)$, $C_5$ a constant chosen to fix the exact proportionality between $b_c(0)$ and $1/Q$, and $b_{min}^{\prime} \equiv b_0/(C_5 Q)$, which cuts $b_T$ off at $O(1/Q)$.  In terms of $\tilde{W}(\T{b},Q,S)$ this modification is to be understood as
\begin{equation} \label{e:Wb}
\tilde{W}(\T{b},Q,S)\to \tilde{W}(\T{b},b_c(b_T),Q,S) \equiv \tilde{W}_{\rm UU}(b_c(b_T),Q) - iM_P\,\epsilon^{ij}b_T^iS_T^j\,\tilde{W}_{\rm UT}^{\rm siv}(b_c(b_T),Q) + \dots\,,
\end{equation}
where
\begin{align}
\tilde{W}_{\rm UU}(b_c(b_T),Q)& = \sum_j H_j(\mu_Q,Q)\,\tilde{f}_1^{j}(x,b_c(b_T);Q^2,\mu_Q)\,\tilde{D}_1^{h/j}(z,b_c(b_T);Q^2,\mu_Q)\,,\label{e:WbUU}\\[0.3cm]
\tilde{W}_{\rm UT}^{\rm siv}(b_c(b_T),Q)& = \sum_j H_j(\mu_Q,Q)\,\tilde{f}_{1T}^{\perp(1)j}(x,b_c(b_T);Q^2,\mu_Q)\,\tilde{D}_1^{h/j}(z,b_c(b_T);Q^2,\mu_Q)\,.\label{e:WbUT}
\end{align}
Note that we have written (\ref{e:Wb}) in such a way that {\it no} kinematic $b_T$ dependence shows up in the scalar functions $\tilde{W}_{\rm UU}(b_c(b_T),Q)$ and $\tilde{W}^{\rm siv}_{\rm UT}(b_c(b_T),Q)$.  That is, the modification $b_T\to b_c(b_T)$ only applies to the parts of $\tilde{W}(\T{b},b_c(b_T),Q,S)$ that  undergo CSS  evolution and {\it not} to any kinematic/tensorial $b_T$  prefactors.  The $b$-space functions in Eqs.~(\ref{e:WbUU}), (\ref{e:WbUT}) are suitably modified to be
\begin{align}
\tilde{f}_1^{j}(x,b_c(b_T);Q^2,\mu_Q) &= \sum_{j'}\int_x^1\!\frac{d\hat{x}} {\hat{x}}\,\tilde{C}^{\rm pdf}_{j/j'}(x/\hat{x},b_*(b_c(b_T));\bar{\mu}^2,\bar{\mu},\alpha_s(\bar{\mu}))\,f_1^{j'}(\hat{x};\bar{\mu})\nonumber\\
&\times\exp\left\{\tilde{K}(b_*(b_c(b_T));\bar{\mu})\ln\left(\frac{Q} {\bar{\mu}}\right)+\int_{\bar{\mu}}^{\mu_Q}\!\frac{d\mu^\prime} {\mu^\prime}\left[\gamma(\alpha_s(\mu^\prime);1)-\ln\left(\frac{Q} {\mu^{\prime}}\right)\gamma_K(\alpha_s(\mu^\prime))\right]\right\}\nonumber\\
&\times\exp\left\{-g_{\rm pdf}(x,b_c(b_T);Q_0,b_{max})-g_K(b_c(b_T);b_{max})\ln\left(\frac{Q}{Q_0}\right)\right\}\,,\label{e:f1}\\[0.3cm]
\tilde{D}_1^{h/j}(z,(b_c(b_T);Q^2,\mu_Q) &= \sum_{i'}\int_z^1\!\frac{d\hat{z}} {\hat{z}^3}\,\tilde{C}^{\rm ff}_{i'/j}(z/\hat{z},b_*(b_c(b_T));\bar{\mu}^2,\bar{\mu},\alpha_s(\bar{\mu}))\,D_1^{h/i'}(\hat{z};\bar{\mu})\nonumber\\
&\times\exp\left\{\tilde{K}(b_*(b_c(b_T));\bar{\mu})\ln\left(\frac{Q} {\bar{\mu}}\right)+\int_{\bar{\mu}}^{\mu_Q}\!\frac{d\mu^\prime} {\mu^\prime}\left[\gamma(\alpha_s(\mu^\prime);1)-\ln\left(\frac{Q} {\mu^{\prime}}\right)\gamma_K(\alpha_s(\mu^\prime))\right]\right\}\nonumber\\
&\times\exp\left\{-g_{\rm ff}(z,b_c(b_T);Q_0,b_{max})-g_K(b_c(b_T);b_{max})\ln\left(\frac{Q}{Q_0}\right)\right\}\,,\label{e:D1}\\[0.3cm]
\tilde{f}_{1T}^{\perp(1)j}(x,b_c(b_T);Q^2,\mu_Q) &= -\frac{1} {2M_P}\sum_{j'}\int_x^1\!\frac{d\hat{x}_1} {\hat{x}_1}\frac{d\hat{x}_2} {\hat{x}_2}\,\tilde{C}^{\rm siv}_{j/j'}(\hat{x}_1,\hat{x}_2,b_*(b_c(b_T));\bar{\mu}^2,\bar{\mu},\alpha_s(\bar{\mu}))\,T_F^{j'}(\hat{x}_1,\hat{x}_2;\bar{\mu})\nonumber\\
&\times\exp\left\{\tilde{K}(b_*(b_c(b_T));\bar{\mu})\ln\left(\frac{Q} {\bar{\mu}}\right)+\int_{\bar{\mu}}^{\mu_Q}\!\frac{d\mu^\prime} {\mu^\prime}\left[\gamma(\alpha_s(\mu^\prime);1)-\ln\left(\frac{Q} {\mu^{\prime}}\right)\gamma_K(\alpha_s(\mu^\prime))\right]\right\}\nonumber\\
&\times\exp\left\{-g_{\rm siv}(x,b_c(b_T);Q_0,b_{max})-g_K(b_c(b_T);b_{max})\ln\left(\frac{Q}{Q_0}\right)\right\}\,,\label{e:Siv}
\end{align}
where
\begin{equation} \label{e:mubar}
b_*(b_c(b_T)) = \sqrt{\frac{b_T^2+b_{min}^{\prime 2}} {1+b_T^2/b_{max}^2+b_{min}^{\prime 2}/b_{max}^2}}\,,\quad\quad\quad \bar{\mu} \equiv \frac{C_1} {b_*(b_c(b_T))}\,,
\end{equation}
with $b'_{min}$ defined after (\ref{e:bcbT}).
\\
\item {\it Define} a new $W$-term,
\begin{align} \label{e:Wnew}
  W(\T{q},Q,S;C_5) &\equiv \Xi(q_T/Q)\int\!\frac{d^2\T{b}} {(2\pi)^2}\,e^{i\T{q}\cdot\T{b}}\,\tilde{W}(\T{b},b_c(b_T),Q,S)\nonumber\\[0.1cm]
&= \Xi(q_T/Q)\left[W_{\rm UU}(q_T,Q;C_5)+|\T{S}|\sin(\phi_h-\phi_S)\,W_{\rm UT}^{\rm siv}(q_T,Q;C_5) + \dots\right],
\end{align}
where
\begin{align}
  W_{\rm UU}(q_T,Q;C_5) &\equiv \int\!\frac{d^2\T{b}} {(2\pi)^2}\,e^{i\T{q}\cdot\T{b}}\,\tilde{W}_{\rm UU}(b_c(b_T),Q)
  =\int_0^\infty \frac{db_T}{2\pi}\, J_0(q_Tb_T)\tilde{W}_{\rm UU}(b_c(b_T),Q)  \,,\label{e:WUU_q}  \\[0.3 cm]
  W_{\rm UT}^{\rm siv}(q_T,Q;C_5) &\equiv -iM_P\int\!\frac{d^2\T{b}} {(2\pi)^2}\,e^{i\T{q}\cdot\T{b}}\,(\boldsymbol{\hat{h}}\cdot\T{b})\,\tilde{W}_{\rm UT}^{\rm siv}(b_c(b_T),Q)=-M_p\int_0^\infty \frac{db_T}{2\pi}b_T^2J_1(q_Tb_T)\tilde{W}_{\rm UT}^{\rm siv}(b_c(b_T),Q)
  \,,\label{e:WUT_q}
\end{align}
with $C_5$ again a constant chosen to optimize the control of large logarithms that arise as $b_T\to 0$.  The quantity $\Xi(q_T/Q)$ in (\ref{e:Wnew}) is a smooth function chosen so that it is unity at $q_T = 0$ and approaches zero for large $q_T \gtrsim Q$~\cite{Collins:2011zzd,Collins:2016hqq}.
This factor ensures that $W(\T{q},Q,S;C_5)$ is sufficiently suppressed for $q_T \gtrsim Q$, where its accuracy has significantly degraded.  
The momentum-space functions are likewise {\it defined} as
\begin{align}
  f_1^{j}(x,k_T;Q^2,\mu_Q;C_5)&\equiv \int_0^\infty \!\frac{db_T}{2\pi}\, b_T J_0(k_Tb_T)\tilde{f}_1^{j}(x,b_c(b_T);Q^2,\mu_Q)
   \,,\label{e:f1kT}\\[0.3cm]
   D_1^{j}(z,zp_T;Q^2,\mu_Q;C_5)&\equiv \int_0^\infty \!\frac{db_T}{2\pi}\, b_T J_0(p_Tb_T)\,\tilde{D}_1^{h/j}(z,b_c(b_T);Q^2,\mu_Q)\,,
   \\[0.3cm]
  \frac{k_T^2} {2M_P^2}\,f_{1T}^{\perp j}(x,k_T;Q^2,\mu_Q;C_5)&\equiv k_T\!\!\int_0^\infty\! \frac{db_T}{4\pi}b_T^2J_1(k_Tb_T)
  \,\tilde{f}_{1T}^{\perp(1)j}(x,b_c(b_T);Q^2,\mu_Q) 
  \,.\label{e:SivkT}
\end{align} 

\item {\it Define} a new $Y$-term,
\begin{equation}
Y(\T{q},Q,S;C_5) \equiv X(q_T/m)\left\{{\rm FO}(\T{q},Q,S)-{\rm AY}(\T{q},Q,S;C_5)\right\}\,,
\end{equation}
where $X(q_T/m)$ is a smooth function that approaches zero for $q_T \lesssim m$ and unity for $q_T \gtrsim m$~\cite{Collins:1984kg,Collins:2016hqq}.  The function $X(q_T/m)$ ensures that $Y(\T{q},Q,S;C_5)$ is sufficiently suppressed for $q_T\lesssim m$, where its accuracy has significantly degraded.
The quantity ${\rm AY}(\T{q},Q,S;C_5)$, being the asymptotic expansion of $W(\T{q},Q,S;C_5)$ at large $q_T$, includes the modifications (I), (II).  The change (II) to ${\rm AY}(\T{q},Q,S)$ has the additional benefit that the integral of the asymptotic term over all $\T{q}$ is now finite, whereas in the original CSS formalism it diverges. We mention that in ${\rm AY}(\T{q},Q,S;C_5)$ the scale $\mubstar$ is replaced with $\mu_Q$ (see, e.g., Ref. [1, Sec. VIII]). Furthermore, as $q_T\to 0$ the singular logarithms cancel between ${\rm FO}(\T{q},Q,S)$ and ${\rm AY}(\T{q},Q,S;C_5)$.  Thus, $Y(\T{q},Q,S;C_5)$ is suppressed for $q_T\ll Q$.
\\ 
\item With these modifications, the $q_T$-differential cross section (\ref{e:SIDIS_WY}) now reads
\begin{equation} \label{e:SIDIS_cs}
\Gamma(\T{q},Q,S) = W(\T{q},Q,S;C_5)+Y(\T{q},Q,S;C_5)+O((m/Q)^{c})\,.
\end{equation}
\end{enumerate}

\subsection{Agreement between TMD and collinear results \label{ss:TMD_Col}}

We start with the cross section in Eq.~(\ref{e:SIDIS_cs}), which can be written as
\begin{subequations}
\begin{align}
\Gamma(\T{q},Q,S) &= \Xi(q_T/Q)\left[W_{\rm UU}(q_T,Q;C_5)+|\T{S}|\sin(\phi_h-\phi_S)\,W_{\rm UT}^{\rm siv}(q_T,Q;C_5) + \dots\right]+Y(\T{q},Q,S;C_5)\\[0.3cm]
&=\left[W_{\rm UU}(q_T,Q;C_5)+|\T{S}|\sin(\phi_h-\phi_S)\,W_{\rm UT}^{\rm siv}(q_T,Q;C_5) + \dots\right]\nonumber\\[0.1cm]
&\;\;\; -\left(1-\Xi(q_T/Q)\right)\left[W_{\rm UU}(q_T,Q;C_5)+|\T{S}|\sin(\phi_h-\phi_S)\,W_{\rm UT}^{\rm siv}(q_T,Q;C_5) + \dots\right] \nonumber\\[0.1cm]
&\;\;\;+ Y(\T{q},Q,S;C_5)\,. \label{e:Gam}
\end{align}
\end{subequations}
Note that at small $q_T\ll Q$, the second and third lines of (\ref{e:Gam}) are suppressed by $q_T/Q$ compared to the first line.  Since they only become sizable for larger $q_T$, the second and third lines contribute at $O(\alpha_s(Q))$.  Therefore, the LO part of any (possibly weighted) $q_T$-integration of $\Gamma(\T{q},Q,S)$ will be from the first line of Eq.~(\ref{e:Gam}).

We now show that the improvements of Sec.~\ref{ss:mod} resolve the problems in the original CSS formalism (see Sec.~\ref{ss:issues}) with integrating $\Gamma(\T{q},Q,S)$, as well as the TMD functions, over transverse momentum.  While $\Xi(q_T/Q)$ and $X(q_T/m)$ in (II), (III) are needed to help accurately describe the intermediate $q_T$ region, as we will see below, it is the $b_T\to b_c(b_T)$ modification of (I) that is crucial to recover the expected relations between TMD and collinear quantities.  For the unpolarized case we find~\cite{Collins:2016hqq}
\begin{align} 
  \frac{d\sigma} {dxdydz} &\equiv 2\int\!d^2 \boldsymbol{P_{h\perp}}\int \!d\phi_S\,\Gamma(\T{q},Q,S)
=4\pi z^2\,\tilde{W}_{\rm UU}^{\rm OPE}(b^\prime_{min},Q)_{\rm LO} + O(\alpha_s(Q)) + O((m/Q)^p)\nonumber\\[0.1cm]
&=\frac{4\pi\alpha_{em}^2} {yQ^2}(1-y+y^2/2)\,\sum_j\,e_j^2\,f_1^{j}(x;\mu_c)\,D_1^{h/j}(z;\mu_c)+O(\alpha_s(Q))+O((m/Q)^p)\,,
\label{e:unpol_parton}
\end{align}
where 
$\mu_c \equiv \lim_{b_T\to 0} \,\bar{\mu}\approx C_1C_5Q/b_0$ (with $\bar{\mu}$ given in (\ref{e:mubar})) so that $\mu_c$ is on the order $Q$.
This agrees with the result in~\cite{Bacchetta:2006tn}.  Note that ``$O(\alpha_s(Q))$'' includes the next-to-leading order (NLO) corrections to the coefficients $\tilde{C}$  and hard factors $H$ along with the terms in the second and third lines of Eq.~(\ref{e:Gam}) (since both are unsuppressed only at large $q_T$), and the $O((m/Q)^p)$ correction is from replacing $\tilde{W}_{\rm UU}(b^\prime_{min},Q)$ with $\tilde{W}_{\rm UU}^{\rm OPE}(b^\prime_{min},Q)$~\cite{Collins:2016hqq}. This result was first derived for the iCSS formalism in Ref.~\cite{Collins:2016hqq}.  

We now extend this to the Sivers case and obtain
\begin{align}
\frac{d\langle P_{h\perp}\,\Delta\sigma(S_T)\rangle} {dxdydz}&\equiv 2\int\!d^2\boldsymbol{P_{h\perp}}\int\!d\phi_S\,P_{h\perp}\sin(\phi_h-\phi_S)\,\Gamma(\T{q},Q,S)\nonumber\\[0.15cm]
&= -4\pi z^3M_P\lim_{b'_T\to 0}\frac{1}{b'_T}\left[\int_0^\infty\! db_T\,b_T\!\int_0^\infty\! dq_T\,q_Tb_T\,J_1(q_Tb'_T)\,J_1(q_Tb_T)\,\tilde{W}_{\rm UT}^{\rm siv}(b_c(b_T),Q)_{\rm LO}\right] + O(\alpha_s(Q))\nonumber\\[0.15cm]
&= -4\pi z^3M_P\lim_{b'_T\to 0}\frac{1}{b'_T}\left[\int_0^\infty\! db_T\,\delta(b_T-b'_T)\,b_T\,\tilde{W}_{\rm UT}^{\rm siv}(b_c(b_T),Q)_{\rm LO}\right] + O(\alpha_s(Q))\nonumber\\[0.15cm]
&=-4\pi z^3M_P\,\tilde{W}_{\rm UT}^{\rm siv,OPE}(b^\prime_{min},Q)_{\rm LO} + O(\alpha_s(Q)) + O((m/Q)^{p'})\nonumber\\[0.1cm]
&=\frac{2\pi\, z\,\alpha_{em}^2} {yQ^2}(1-y+y^2/2)\!\sum_j\!e_j^2\, T_F^{j}(x,x;\mu_c)\,D_1^{h/j}(z;\mu_c)+O(\alpha_s(Q))+O((m/Q)^{p'})\,.\label{e:Siv_parton}
\end{align}
Again we confirm the previous LO calculations in the literature~\cite{Kang:2012ns}.  Note as before that ``$O(\alpha_s(Q))$'' includes the NLO corrections to the coefficients $\tilde{C}$ and hard factors $H$ along with the terms in the second and third lines of Eq.~(\ref{e:Gam}), and the $O((m/Q)^{p'})$ correction is from replacing $\tilde{W}^{\rm siv}_{\rm UT}(b^\prime_{min},Q)$ with $\tilde{W}_{\rm UT}^{\rm siv,OPE}(b^\prime_{min},Q)$.  Again in going from the second to the third line have used $\int_0^\infty\! dq_T\,q_TJ_n(q_Tb'_T)\,J_m(q_Tb_T) = \delta_{nm}\,\delta(b_T-b'_T)/b_T$.  

We emphasize that it was crucial in (\ref{e:WUT_q}) that the $b_T$ in $(\boldsymbol{\hat{h}}\cdot\T{b})$ {\it not} get replaced by $b_c(b_T)$ in order to achieve the result (\ref{e:Siv_parton}).  This manifests itself in the second line of (\ref{e:Siv_parton}), where the factor $(q_T b_T)$ appears instead of $(q_T b_c(b_T))$.  If, on the other hand, the $b_T\to b_c(b_T)$ replacement was made in $(\boldsymbol{\hat{h}}\cdot\T{b})$, the third line in (\ref{e:Siv_parton}) would give a divergent result since then one would have a factor $\lim_{b'_T\to 0}b_c(b'_T)/b'_T = \lim_{b'_T\to 0}b'_{min}/b'_T$.  This example highlights the key observation needed in order to use the iCSS formalism with polarized observables.  In general it is a statement that the $b_T\to b_c(b_T)$ prescription only applies to the $b_T$ dependence that is a part of the evolution and {\it not} to any external (kinematic) $b_T$ prefactors.

In terms of the momentum-space functions (\ref{e:f1kT})--(\ref{e:SivkT}), we also find 
\begin{align}
\int\!d^2\T{k}\,f_1^{j}(x,k_T;Q^2,\mu_Q;C_5)&=\tilde{f}_1^{j}(x,b^\prime_{min};Q^2,\mu_Q)=f_1^{j}(x;\mu_c) + O(\alpha_s(Q)) + O((m/Q)^p)\label{e:f1_col}\,,\\[0.3cm]
z^2\!\int\!d^2\T{p}\,D_1^{j}(z,zp_T;Q^2,\mu_Q;C_5)&=z^2\tilde{D}_1^{h/j}(z,b^\prime_{min};Q^2,\mu_Q)=D_1^{h/j}(z;\mu_c) + O(\alpha_s(Q)) + O((m/Q)^p)\label{e:D1_col}\,,\\[0.3cm]
\int\!d^2\T{k}\,\frac{k_T^2} {2M_P^2}\,f_{1T}^{\perp j}(x,k_T;Q^2,\mu_Q;C_5)&\equiv f_{1T}^{\perp(1)\,j}(x;Q^2,\mu_Q; C_5)\nonumber\\
&=\tilde{f}_{1T}^{\perp(1)j}(x,b^\prime_{min};Q^2,\mu_Q)=-\frac{1} {2M_P}T^j_F(x,x;\mu_c)+ O(\alpha_s(Q)) + O((m/Q)^{p'})\label{e:Siv_col}\,,
\end{align}
where again
$\mu_c \equiv \lim_{b_T\to 0} \,\bar{\mu}\approx C_1C_5Q/b_0$ (with $\bar{\mu}$ given in (\ref{e:mubar})) so that $\mu_c$ is on the order $Q$.\footnote{Phenomenological fits of TMDs use $C_1$ and $b_{max}$ to optimize the perturbation theory.  The fact that the collinear functions on the r.h.s.~of (\ref{e:f1_col})--(\ref{e:Siv_col}) depend on these parameters via $\mu_c$ is a result of the truncation of the perturbative series in $\alpha_s$.}  Note that, due to the $b_T \to b_c(b_T)$ modification, the above integrals on the l.h.s.~are UV finite, yielding at LO and for $Q\gg m$ the renormalized collinear functions on the far r.h.s. To obtain these last equalities we have used the fact that $b^\prime_{min}\sim O(1/Q)$ so that we can replace (\ref{e:f1})--(\ref{e:Siv}) with their OPE pieces and expand the exponentials in powers of $\alpha_s(Q)$ without large logarithms.  The correction terms also include NLO in the coefficients $\tilde{C}$.  The results in Eqs.~(\ref{e:f1_col})--(\ref{e:Siv_col}) agree with our expectations from the ``na\"{i}ve'' operator definitions of TMDs.\footnote{We will discuss specifically what we mean by ``na\"{i}ve'' operator definition in Sec.~\ref{s:physical}.}  In particular, Eq.~(\ref{e:Siv_col}) is the well-known relation between the first $k_T$-moment of the Sivers function and the Qiu-Sterman function~\cite{Boer:2003cm} (see also \cite{Meissner:2007rx,Kang:2011mr,Kang:2011hk,Boer:2014bya}). We emphasize that the relations between the integrals of the TMDs on the far l.h.s.~of Eqs.~(\ref{e:f1_col})--(\ref{e:Siv_col}) and the functions $\tilde{f}_1(x,b^\prime_{min};Q^2,\mu_Q)$, $\tilde{D}_1(z,b^\prime_{min};Q^2,\mu_Q)$, and $\tilde{f}_{1T}^{\perp(1)}(x,b^\prime_{min};Q^2,\mu_Q)$, respectively, which can be obtained from Eqs.~(\ref{e:f1})--(\ref{e:Siv}) by setting $b_T$ to zero, hold to all orders in the strong coupling.

\subsection{Power counting in the region $m\ll q_T\ll Q$ \label{ss:power}}
A necessary condition for a TMD modulation in $W(\T{q},Q,S;C_5)$ to yield the corresponding LO collinear cross section (upon a suitably weighted integration over $\T{q}$) is that there must be no power-counting mismatch in the intermediate $q_T$ region ($m \ll q_T \ll Q$) where both factorization in terms of TMD functions and collinear functions are valid.\footnote{Most likely this is also a sufficient condition at LO.  However, at NLO, a quantitative matching in the intermediate-$q_T$ region is probably also required in order for the integrated TMD results to match the collinear ones at that order in $\alpha_s$.}  For the Sivers effect that we focused on in Sec.~\ref{ss:TMD_Col}, it was shown explicitly in SIDIS~\cite{Ji:2006br,Koike:2007dg} and Drell-Yan~\cite{Ji:2006ub} that the results for TMD and collinear twist-3 factorization match in the $m \ll q_T \ll Q$ regime.  The same was also explicitly proven for the Collins effect in SIDIS~\cite{Yuan:2009dw}, for hyperon production in SIDIS involving the Boer-Mulders function~\cite{Zhou:2008fb}, and for certain modulations in Drell-Yan involving the worm-gear functions~\cite{Zhou:2009jm}.  In addition, this problem was discussed extensively in Ref.~\cite{Bacchetta:2008xw}, where the authors analyzed if the powers of $q_T$ matched in the $m \ll q_T \ll Q$ region between results at low and high transverse momentum for the structure functions that enter the SIDIS cross section.  
For those structure functions where calculations existed in the literature for both regions, there was only a mismatch at intermediate $q_T$ for the $\cos(2\phi_h)$ and $\sin(3\phi_h-\phi_S)$ modulations.  

\subsection{Other polarized observables \label{ss:other}}
Our extension of the iCSS procedure to the Sivers effect is quite general and can also be used for any polarized observable.  Here we outline the basic steps.  First, one would continue with the expansion in Eq.~(\ref{e:Wb}) to include other polarized terms.  We generically denote these by $C^{\rm pol}(M_P,\T{b},S)\,\tilde{W}^{\rm pol}(b_c(b_T),Q)$, where $\tilde{W}^{\rm pol}(b_c(b_T),Q)$ is a scalar function (initially of $b_T$ before the $b_T \to b_c(b_T)$ modification), and $C^{\rm pol}(M_P,\T{b},S)$ is the associated tensor structure.  The key is that {\it all} kinematic $b_T$ dependence must be contained in $C^{\rm pol}(M_P,\T{b},S)$.  One can receive guidance for the structure of these terms from Eqs.~(2.13), (2.14) of Ref.~\cite{Boer:2011xd}, where, exactly like in (\ref{e:f1})--(\ref{e:Siv}), the evolution of the $b$-space functions that enter those formulas does not contain any kinematic $b_T$ prefactors.  Next, one would continue with the modifications (II) and (III), where, in particular, one defines the momentum-space quantities in terms of the modified $b$-space ones, exactly like in Eqs.~(\ref{e:Wnew})--(\ref{e:WUT_q}) and Eqs.~(\ref{e:f1kT})--(\ref{e:SivkT}).  Finally, one puts these pieces together in the $q_T$-differential cross section of (IV).  Given our discussion in Sec.~\ref{ss:power}, we expect that all TMD functions one obtains from the iCSS procedure will reduce at LO to their collinear counterparts upon integration (like in (\ref{e:f1_col})--(\ref{e:Siv_col})) while inconsistencies could arise when integrating certain modulations in $W(\T{q},Q,S;C_5)$, specifically $\cos(2\phi_h)$ and $\sin(3\phi_h-\phi_S)$)~\cite{Bacchetta:2008xw}, at LO to obtain the corresponding collinear cross sections.

\section{Physical interpretation of Eqs.~(\ref{e:f1_col})--(\ref{e:Siv_col}) \label{s:physical}}
An important consequence of Eqs.~(\ref{e:f1_col})--(\ref{e:Siv_col}) is that the ``na\"{i}ve'' operator definition interpretation of TMDs is restored at LO.  For example, one can determine the average transverse momentum of unpolarized quarks in a transversely polarized spin-$\frac{1} {2}$ target according to~\cite{Boer:1997bw,Boer:1997nt,Burkardt:2003yg,Burkardt:2004ur,Meissner:2007rx}
\begin{align} \label{e:avgTM}
\langle k_T^i (x;\mu) \rangle_{UT} &=  \frac{1} {2} \int d^2\T{k} \,k_T^i\int\frac{db^-}{2\pi}\int\frac{d^2\T{b}} {(2\pi)^2}\,e^{ixP^+b^-}e^{-i\T{k}\cdot\T{b}}\,\langle P,S|\bar{\psi}(0)\gamma^+\mathcal{W}_{\rm DIS}(0;b)\psi(b)| P,S\rangle\bigg |_{b^+=0}\nonumber \\
&=\frac{1} {2}\int\!\frac{db^- dy^-} {4\pi}\,e^{ixP^+b^-}\langle P,S|\bar{\psi}(0)\gamma^+\mathcal{W}(0;y^-)gF^{+i}(y^-)\,\mathcal{W}(y^-;b^-)\psi(b^-)|P,S\rangle\nonumber\\
&=\frac{1} {2}\epsilon^{ij}S_T^j\,T_F(x,x;\mu)\,,
\end{align}
where $\mathcal{W}_{\rm DIS}(0;b)$ is a future-pointing staple gauge link connecting $(0^+,0^-,\T{0})$ to $(0^+,b^-,\T{b})$, and we have used (\ref{e:T_F}) in going from the second to the last line.  (The relation (\ref{e:avgTM}) holds for each quark flavor.) For Drell-Yan, where one uses a past-pointing staple gauge link, there will be a sign change on the r.h.s.~of Eq.~(\ref{e:avgTM})~\cite{Collins:2002kn}.  That is, $\langle k_T^i (x;\mu) \rangle_{UT}$ is process-dependent. Note that the operator defining the TMD $f_{1T}^{\perp j}(x,k_T;Q^2,\mu_Q;C_5)$ on the l.h.s.~of (\ref{e:Siv_col}) includes a UV renormalization factor and a soft factor, along with non-light-like Wilson lines (see, e.g., Refs.~\cite{Collins:2011zzd,Aybat:2011ge}).  However, this is {\it not} the operator that enters the first line of Eq.~(\ref{e:avgTM}).  Rather, the TMD operator that sits in (\ref{e:avgTM}) is the ``na\"{i}ve'' definition, where the UV renormalization and soft factors are kept to LO and the Wilson lines are on the lightcone.  On the other hand, the collinear operator in the second line of Eq.~(\ref{e:avgTM}) {\it is} the one which underlies the Qiu-Sterman function $T_F(x,x;\mu_c)$ on the r.h.s.~of (\ref{e:Siv_col}).\footnote{The scale dependence is from the renormalized correlator one defines since the $k_T$-integration in the first line of (\ref{e:avgTM}) is UV divergent.}  Therefore, it is the Qiu-Sterman function which fundamentally is related to average transverse momentum, and, due to Eq.~(\ref{e:Siv_col}), the first $k_T$-moment of the Sivers function (within the iCSS formalism) retains this interpretation at LO.  We mention that both the l.h.s.~and the r.h.s.~of Eq.~(\ref{e:avgTM}), i.e., the operators in the first line and second line, respectively, are implicitly renormalized using the same procedure.  Thus, both functions are modified in the same way from the strict physical interpretation one obtains from using bare operators.

Note that the interpretation of the Qiu-Sterman function given in (\ref{e:avgTM}) is compatible with the understanding of the average transverse force acting on quarks in a transversely polarized spin-$\frac{1} {2}$ target~\cite{Burkardt:2008ps}.  Moreover, relation (\ref{e:Siv_col}) made it possible to connect TSSAs in different processes (e.g., the Sivers effect in SIDIS and $A_N$ in proton-proton collisions) and has been used routinely in phenomenology (see, e.g., Refs.~\cite{Metz:2012ui,Kang:2012xf,Gamberg:2013kla,Kanazawa:2014dca,Kanazawa:2014nea}).  The incorporation of evolution in the TMD correlator through the original CSS formalism breaks the ``na\"{i}ve'' relations between TMDs and collinear functions (see Eqs.~(\ref{e:f1_col_old})--(\ref{e:Siv_col_old})).  Nevertheless, as we have shown above, the modifications implemented by the iCSS framework allow one to preserve these identities at LO.

\section{Summary \label{s:sum}} 
In this Letter we have extended the improved CSS formalism of Ref.~\cite{Collins:2016hqq} to the case of polarized observables, which are especially important for experiments studying the 3D structure of hadrons.  As a result, we are able to recover the well-known relations between TMD and collinear quantities one expects from their na\"{i}ve operator definitions.  For example, we have shown at LO that the weighted Sivers effect, using the iCSS $W+Y$ construction of TMD factorization, yields the collinear twist-3 result in the literature. We also have demonstrated the validity of the relation between the first $k_T$-moment of the Sivers function and the Qiu-Sterman function, which holds at LO in iCSS.  Since the latter fundamentally defines the average transverse momentum of unpolarized quarks inside a transversely polarized spin-$\frac{1}{2}$-target (see Eq.~(\ref{e:avgTM})), the former retains this interpretation as well at LO.  We have discussed how the iCSS modifications can be applied to other polarized observables, where one would also recover the other known identities between TMD and collinear twist-3 functions.  We leave as future work the implementation of the iCSS method into phenomenological analyses.
 
\section*{Acknowledgments}
The authors thank J.~Collins for a careful reading of the manuscript, and D.~Boer, J.-w.~Qiu, T.~Rogers, N.~Sato, and W.~Vogelsang for many fruitful discussions on this topic.  L.G. thanks Jefferson Lab for  support during  sabbatical leave, where part of this work was completed. This material is based upon work supported by the
U.S. Department of Energy, Office of Science, Office of Nuclear
Physics under Award No. DE-FG02-07ER41460 (L.G.),
No.~DE-AC05-06OR23177 (A.P.),  by the National Science Foundation 
under Contract No. PHY-1516088 (A.M.), 
No.~PHY-1623454 (A.P.), and within the 
framework of the TMD Topical Collaboration.


\end{document}